\documentclass[11pt,a4paper]{article}
\usepackage[hyperref]{emnlp2020}
\usepackage{times}
\usepackage{latexsym}

\usepackage{enumitem}

\usepackage{microtype}

\usepackage{booktabs}       
\usepackage{amsfonts}       
\usepackage{nicefrac}       
\usepackage{microtype}      
\usepackage{lipsum}
\usepackage{graphicx}
\usepackage{todonotes}
\usepackage{xspace}
\usepackage[hang,flushmargin]{footmisc}

\newcommand{\sys}{\textsc{Sledge}\xspace}

\aclfinalcopy 

\title{SLEDGE: A Simple Yet Effective Baseline for \\ COVID-19 \underline{S}cientific Know\underline{ledge} Search}

\author{Sean MacAvaney$^\dag$ \hspace{1.7em}
Arman Cohan$^\ddag$ \hspace{1.7em}
{\bf Nazli Goharian$^\dag$} \hspace{1.4em}
 \vspace{6pt}\\
  $^\dag$Information Retrieval Lab,
  Georgetown University, Washington DC  \vspace{2pt} \\
  $^\ddag$Allen Institute for AI, Seattle WA \vspace{2pt}\\
  {\tt \{sean,nazli\}@ir.cs.georgetown.edu, armanc@allenai.org}
}

\date{}

\begin{document}
\maketitle
\begin{abstract}
With worldwide concerns surrounding the Severe Acute Respiratory Syndrome Coronavirus 2 (SARS-CoV-2), there is a rapidly growing body of literature on the virus. Clinicians, researchers, and policy-makers need a way to effectively search these articles. In this work, we present a search system called \sys, which utilizes SciBERT to effectively re-rank articles. We train the model on a general-domain answer ranking dataset, and transfer the relevance signals to SARS-CoV-2 for evaluation. We observe \sys's effectiveness as a strong baseline on the TREC-COVID challenge (topping the learderboard with an nDCG@10 of 0.6844). Insights provided by a detailed analysis provide some potential future directions to explore, including the importance of filtering by date and the potential of neural methods that rely more heavily on count signals. We release the code to facilitate future work on this critical task.\footnote{ \url{https://github.com/Georgetown-IR-Lab/covid-neural-ir}}
\end{abstract}


\section{Introduction}

The emergence of the Severe Acute Respiratory Syndrome Coronavirus 2 (SARS-CoV-2) prompted a worldwide research response. In the first 100 days of 2020, over 5,000 research articles were published related to SARS-CoV-2 or COVID-19. Together with articles about similar viruses researched before 2020, the body of research exceeds 50,000 articles.  This results in a considerable burden for those seeking information about various facets of the virus, including researchers, clinicians, and policy-makers.

In the interest of establishing a strong baseline for retrieving scientific literature related to COVID-19, we introduce \sys: a simple yet effective baseline for coronavirus \underline{S}cientific know\underline{LEDGE} search.
Our baseline utilizes a combination of state-of-the-art techniques for neural information retrieval.

Recent work in neural information retrieval shows the effectiveness of pretrained language models in document ranking. \cite{MacAvaney2019CEDRCE,Nogueira2019PassageRW,Hofsttter2020InterpretableT,Dai2019DeeperTU,nogueira2020document}. Building upon success of these models, \sys is comprised of a re-ranker based on SciBERT \cite{Beltagy2019SciBERTPC}, a pretrained language model optimized for scientific text. Since at the time of writing there is no available training data for COVID-19 related search, we additionally use a domain transfer approach by training \sys on MS-MARCO \cite{Campos2016MSMA}, a general-domain passage ranking dataset, and apply it to COVID-19 literature search in zero-shot setting. 

We show that \sys achieves strong results in the task of scientific literature search related to COVID-19. In particular, \sys tops the leaderboard in Round 1 of the TREC-COVID Information Retrieval shared task~\cite{trec-covid},\footnote{\url{https://ir.nist.gov/covidSubmit/}} a new test bed for evaluating effectiveness of search methods for COVID-19. We also provide an analysis into the hyperparameter tuning conducted, the effect of various query and document fields, and possible shortcomings of the approach. Insights from the analysis highlight the importance of a date filter for improving precision, and the possible benefit of utilizing models that include count-based signals in future work. We hope that better natural language processing and search tools can contribute to the fight against the current global crisis.

\vspace{1em}

\section{Related Work}

Retrieval of scientific literature has been long-studied~\cite{Lawrence1999IndexingAR,Lalmas2007INEX2,hersh2009trec,Lin2008IsSF,medlar2016pulp,sorkhei2017exploring,huang2019holes}. Most recent work for scientific literature retrieval has focused on tasks such as collaborative filtering~\cite{Chen2018ResearchPR}, citation recommendation~\cite{Nogueira2020EvaluatingPT}, and clinical decision support~\cite{Soldaini2017LearningTR}, rather than ad-hoc retrieval.

Pre-trained neural language models (such as BERT~\cite{Devlin2019BERTPO}) have recently shown to be effective when fine-tuned for ad-hoc ranking. \citet{Nogueira2019PassageRW} demonstrate that these networks can be fine-tuned for passage ranking tasks. Others later observed effectiveness at document ranking tasks, showing that these models can handle natural-language questions better than prior approaches~\cite{Dai2019DeeperTU} and that they can be incorporated into prior neutral ranking techniques~\cite{MacAvaney2019CEDRCE}.
Although computationally expensive, researches have shown that this can be mitigated to an extent by employing more efficient modeling choices~\cite{Hofsttter2020InterpretableT,MacAvaney2020ExpansionVP}, caching intermediate representations~\cite{Khattab2020ColBERTEA,MacAvaney2020EfficientDR,Gao2020EARLST}, or by modifying the index with new terms or weights~\cite{Nogueira2019DocumentEB,Dai2019ContextAwareST,Nogueira2019FromDT}.
These models also facilitate effective relevance signal transfer; \citet{Yilmaz2019CrossDomainMO} demonstrate that the relevance signals learned from BERT can easily transfer across collections (reducing the chance of overfitting a particular collection).
In this work, we utilize relevance signal transfer from an open-domain question answering dataset to the collection of COVID-19 scientific literature.

In terms of biomedical-related ranking, \citet{MacAvaney2020RankingSD} observed the importance of using a domain-tuned language model (SciBERT~\cite{Beltagy2019SciBERTPC}) when ranking in the biomedical domain (albeit working with clinical text rather than scientific literature). Some work already investigates document ranking and Question Answering (QA) about COVID-19. \citet{Zhang2020RapidlyDA} chronicled their efforts of building and deploying a search engine for COVID-19 articles, utilizing a variety of available tools ranking techniques. In this work, we find that our approach outperforms this system in terms of ranking effectiveness. \citet{Tang2020RapidlyBA} provide a QA dataset consisting of 124 COVID-19 question-answer pairs.

\section{\sys}

This section describes the details of \sys, our method for searching scientific literature related to COVID-19.
We utilize a standard two-stage re-ranking pipeline for retrieving and ranking COVID-19 articles. The articles are curated from the CORD19 dataset \cite{Wang2020CORD19TC} and provided by the task organizers. The first stage employs an inexpensive ranking model (namely, BM25) to generate a high-recall collection of candidate documents. The second stage re-ranks the candidate documents using an expensive but high-precision SciBERT-based~\cite{Beltagy2019SciBERTPC} neural ranking model.

\subsection{First-Stage Retrieval}

We first index the document collection using standard pre-processing methods: English stopword removal and Porter stemming. For the text, we use a concatenation of the title, abstract, and fulltext paragraphs and fulltext headings. The fulltext gives more opportunities for the first-stage ranker to match potentially relevant documents than the title alone would provide. When both the PDF and PubMed XML versions are available, we use the text extracted from the PubMed XML because it is generally cleaner. We then query the index for each topic with BM25. In this system, we used a fixed re-ranking threshold of 500; thus only the top 500 BM25 results are retrieved. In our experiments, we found that there was little recall gained beyond 500.

\subsection{Neural Re-Ranking}

To best capture the domain-specific language related to scientific text we use the SciBERT~\cite{Beltagy2019SciBERTPC} pretrained language model as the basis of a second-stage supervised re-ranker. This model is akin to the Vanilla BERT ranker from~\cite{MacAvaney2019CEDRCE}, but utilizing the SciBERT model base (which is trained on scientific literature) instead. The query and document text are encoded sequentially, and relevance prediction is calculated based on the \texttt{[CLS]} token's representation (which was used for next sentence prediction during pre-training). Documents longer than the maximum length imposed by the positional embeddings are split into arbitrary equal-sized passages. We refer the reader to~\cite{MacAvaney2019CEDRCE} for more details about Vanilla BERT.

At the time of writing there is no training data available for the COVID-19 related search and collecting such data is expensive.
To mitigate this challenge, we utilize a domain transfer approach and apply the learned model to the new domain in a zero-shot setting. This approach also has the advantage of avoiding overfitting on the target dataset. Specifically, we train our model using the standard training sequence of the MS-MARCO passage ranking dataset~\cite{Campos2016MSMA}. This dataset consists of over 800,000 query-document pairs in the general domain with a shallow labeling scheme (typically fewer than two positive relevance labels per query; non-relevance assumed from unlabeled passages). During model training, we employ the following cross-entropy loss function from~\citet{Nogueira2019PassageRW}:
\begin{equation}\small
L(q,d^+,d^-)=-\log(R(q,d^+))-\log(R(q,d^-))
\end{equation}
where $\mathit{q}$ is the query, $\mathit{d^+}$ and $\mathit{d^-}$ are the relevant and non-relevant training documents, and $R(q,d)$ is the relevance score.

\section{Experimental setup}\label{sec:exp}

We now explore the ranking effectiveness of our approach. We evaluate the performance of \sys using the TREC-COVID Information Retrieval Challenge dataset (round 1) \cite{trec-covid}. TREC-COVID uses the CORD-19 document collection~\cite{Wang2020CORD19TC} (2020-04-10 version, 51,045 articles), with a set of 30 topics related to COVID-19. These topics include natural queries such as: \textit{Coronavirus
response to weather changes} and \textit{Coronavirus social distancing impact}. The top articles of participating systems (56 teams) were judged by expert assessors, who rated each article non-relevant (0), partially-relevant (1), or fully-relevant (2) to the topic. In total, 8,691 relevance judgments were collected, with 74\% non-relevant, 13\% partially relevant, and 14\% fully-relevant.

Since the relevance judgments in this dataset are shallow (avg. 290 per query), we measure effectiveness of each system using normalized Discounted Cumulative Gain with a cutoff of 10 (nDCG@10), Precision at 5 of partially and fully-relevant documents (P@5), and Precision at 5 of only fully relevant documents (P@5 (Rel.)). Both nDCG@10 and P@5 are official task metrics; we include the P@5 filtered to only fully-relevance documents because it exposed some interesting trends in our analysis. Since not all submissions contributed to the judgment pool, we also report the percentage of the top 5 documents for each query that have relevance judgments (judged@5). These settings represent a high-precision evaluation; we leave it to future work to evaluate techniques for maximizing system recall, which may require special considerations~\cite{Grossman2015TREC2T}.

Our work utilizes a variety of existing open-source tools, including OpenNIR~\cite{macavaney:wsdm2020-onir}, Anserini~\cite{Yang2017AnseriniET}, and the HuggingFace Transformers library~\cite{Wolf2019HuggingFacesTS}. Our experiments were conducted with a Quadro RTX 8000 GPU, and a learning rate of $2\times10^{-5}$.

\paragraph{Note on manual vs automatic runs} TREC-COVID makes the distinction between manual and automatic runs. We adhere to the broad definition of manual runs, as specified by the task guidelines: ``Automatic construction is when there is no human involvement of any sort in the query construction process; manual construction is everything else... If you make any change to your retrieval system based on the content of the TREC-COVID topics (say add words to a dictionary or modify a routine after looking at retrieved results), then your runs are manual runs.''\footnote{\url{https://ir.nist.gov/covidSubmit/round1.html}} In short, making any change to the system on the basis of observations of the query and/or results qualify as a manual run.

\begin{table*}
\centering
\small
\scalebox{0.97}{
\begin{tabular}{@{}lrrrrl@{}}
\toprule
System & nDCG@10 & P@5 & P@5 (Rel.) & judged@5 & Human intervention \\
\midrule
SLEDGE (ours, ``run1'')           & \bf 0.6844 & 0.7933 & \bf 0.6533 & 100\% & Hyperparameter tuning on subset of queries \\
BBGhelani2              & 0.6689 & 0.8200 & 0.5600 & 100\% & Human-in-loop active learning \\
xj4wang\_run1           & 0.6513 & \bf 0.8333 & 0.5933 & 100\% & Human-in-loop active learning \\
UIUC\_DMG\_setrank\_ret & 0.6082 & 0.7133 & 0.5333 & 100\% & \textit{unspecified} \\
OHSU\_RUN2              & * 0.5966 & 0.7000 & * 0.5200 & 100\% & Query reformulation \& hyperparameter tuning \\
cu\_dbmi\_bm25\_1       & * 0.5887 & 0.7200 &  0.5667 & 96\% & Query reformulation \\
sheikh\_bm25\_manual    & * 0.5790 & 0.7267 & * 0.5333 & 93\% & Query reformulation \\
crowd1                  & * 0.5571 & 0.7067 & * 0.4933 & 93\% & Manual relevance feedback \\
CSIROmed\_RF            & * 0.5479 & * 0.6400 & * 0.5267 & 86\% & Manual relevance feedback \\
dmis-rnd1-run3         & * 0.4649 & * 0.5867 & * 0.4733 & 100\% & Query reformulation \\
\bottomrule
\end{tabular}
}
\caption{Top results using any human intervention (manual runs). * indicates our system exhibits a statistically significant improvement (paired t-test, $p<0.05$).}
\label{tab:manual}
\end{table*}
\begin{table*}
\centering
\small
\begin{tabular}{@{}lrrrrl@{}}
\toprule
 System & nDCG@10 & P@5 & P@5 (Rel.) & judged@5 & Methodology \\
\midrule
 sab20.1.meta.docs & \bf 0.6080 & \bf 0.7800 & 0.4867 & 100\% & VSM, Multi-Index, Lnu.ltu weighting\\ 
 SLEDGE (ours, ``run2'') & 0.6032 & 0.6867 & 0.5667 & 88\% & BM25 + SciBERT \\
 IRIT\_marked\_base & 0.5880 & 0.7200 & 0.5400 & 100\% & BM25+RM3 + BERT-base \\
 CSIROmedNIR* & 0.5875 & 0.6600 & \bf 0.5867 & 76\% & CovidBert-nli + Inv. Index  \\
 base.unipd.it & 0.5720 & 0.7267 & 0.5200 & 95\% & Elastic search + boolean queries\\
 udel\_fang\_run3 & 0.5370 & 0.6333 & 0.4267 & 98\% & F2EXP + CombSUM \\
 uogTrDPH\_QE & 0.5338 & 0.6400 & 0.4667 & 100\% & Terrier + Query Exp. \\
 UP-rrf5rnd1 & 0.5316 & 0.6800 & 0.4800 & 100\% &  unsupervised reciprocal rank fusion \\
 BioinfoUA-emb & 0.5298 & 0.6333 & 0.4733 & 100\% & BM25 + DeepRank traind on BioAsk \\
 UIowaS\_Run3 & 0.5286 & 0.6467 & 0.4733 & 100\% & BM25 + filtering \\
\bottomrule
\end{tabular}
\caption{Top results without using any human intervention (automatic runs). No results exhibit a statistically significant difference compared to our system (paired t-test, $p<0.05$). "*" indicates that some sort of manually specified filtering was used which may contradict the definition of an automatic run by TREC (see note in Section~\ref{sec:exp}).}
\label{tab:auto}
\end{table*}

\section{Results}\label{sec:res}

In this section we discuss our results in two evaluation settings. In the first setting we apply light hyperparmeter tuning on the pipeline which still counts as a \textit{manual} run as discussed in \S\ref{sec:exp}. In the second setting we do not perform any tuning of any sort and thus this setting is an \textit{automatic} run.

\subsection{Ranking with light hyperparmeter tuning}\label{sec:manual}

Recall that the first stage of \sys is based on an initial BM25 ranker, topics in the TREC Covid dataset include 3 different fields: query, question and narrative, and the documents have title, abstract and full-text. Choices of the BM25 parameters and which fields to include in the pipeline can affect the final performance.
Therefore, in the first setting, we lightly tune these hyperparmeters using minimal human judgments on a subset of the topics. Specifically, we use shallow relevance judgments from 15 out of 30 topics assessed by non-experts.\footnote{Topics 1, 2, 6, 8, 9, 11, 13, 17, 18, 20, 21, 24, 27, 29, 30. 849 judgments were made in total. We found that our non-expert annotations did not align well with the officially released expert annotations \S\ref{sec:hypertune}.} Unlike manual runs that require human intervention for query reformulation, active learning, or relevance feedback, we expect our system to be able to generalize to unseen queries in the domain because we use manual relevance signals only for hyperparameter tuning. By tuning the hyperparmeters of the initial retrieval method, the fields of the topic (query, question, narrative) and document text (title, abstract, full text), and a date filter, we found the following pipeline to be most effective based on our non-expert annotations (run tag: run1):
\begin{enumerate}
\item Initial retrieval using BM25 tuned for recall using a grid search ($k_1=3.9$, $b=0.55$), utilizing the keyword query field over the full text of the article. Articles from before January 1, 2020 are disregarded. 
\item Re-ranking using a Vanilla SciBERT model trained on MS-MARCO. The topic's question field is scored over the article's title and abstract.
\end{enumerate}

We report the performance of the top system from the top 10 teams (among manual runs) for TREC-COVID in Table~\ref{tab:manual}. Since the utilization of humans-in-the-loop vary considerably, we also indicate for each run the reported human intervention. We find that \sys outperforms all the other manual runs in terms of nDCG@10 and P@5 (relevant only). Of the top 10 systems that report their technique for human intervention, ours is also the only one that relies on human judgments solely for hyperparameter tuning. This is particularly impressive because the next best systems (BBGhelani2 and xj4wang\_run1) involves human-in-the-loop active learning to rank documents based on the manual assessor's relevance. In terms of statistical significance (paired t-test, $p<0.05$), our approach is on par with these active learning runs, and better than most other submissions in terms of nDCG@10 and P@5 (relevant).

\subsection{Ranking without hyperparameter tuning}\label{sec:auto}

\begin{figure*}
\centering
(a)\hspace{1.85in}(b)\hspace{1.9in}(c)\hspace{0.15in}~
\\
\includegraphics[scale=0.315]{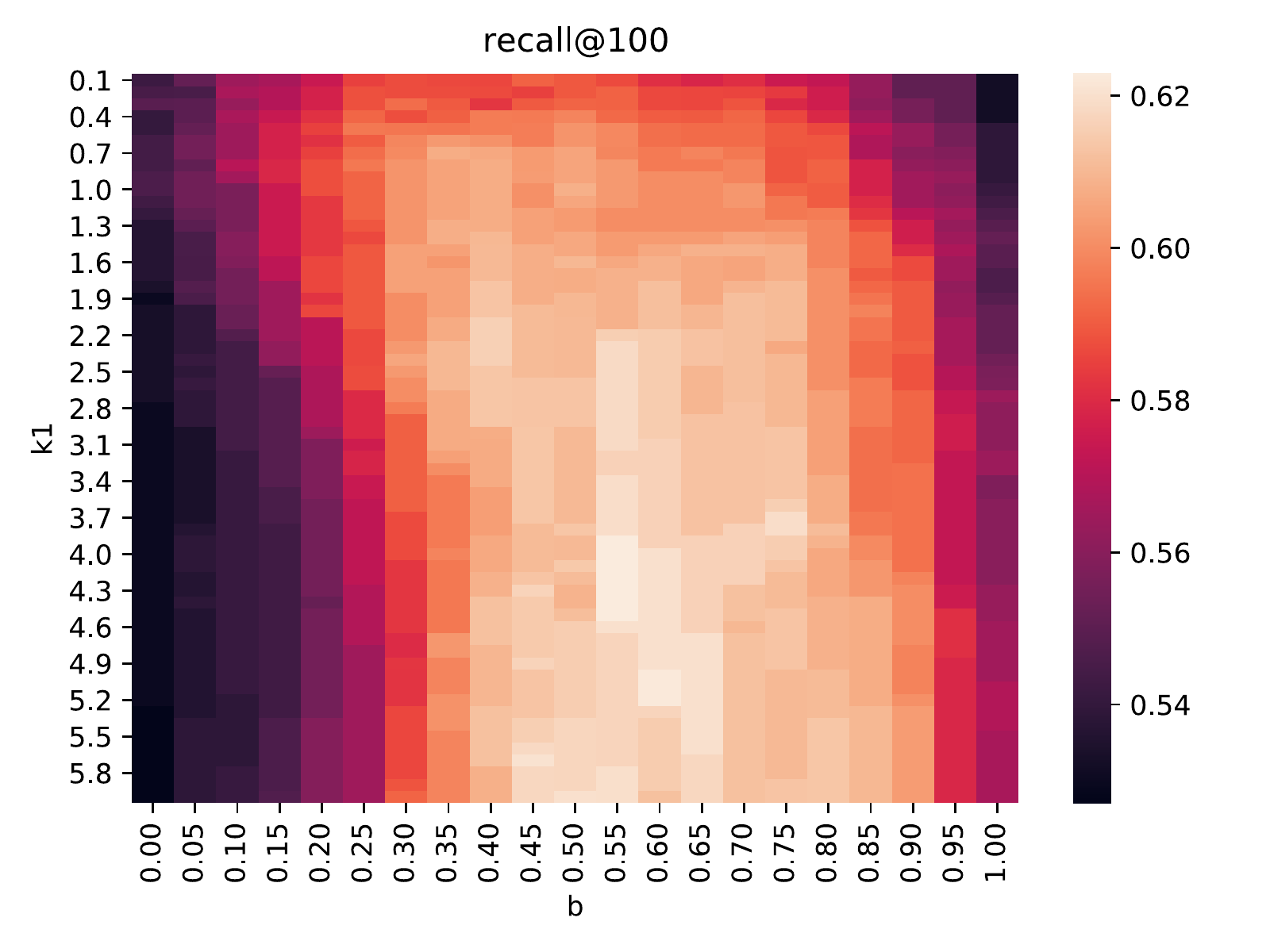}
\includegraphics[scale=0.315]{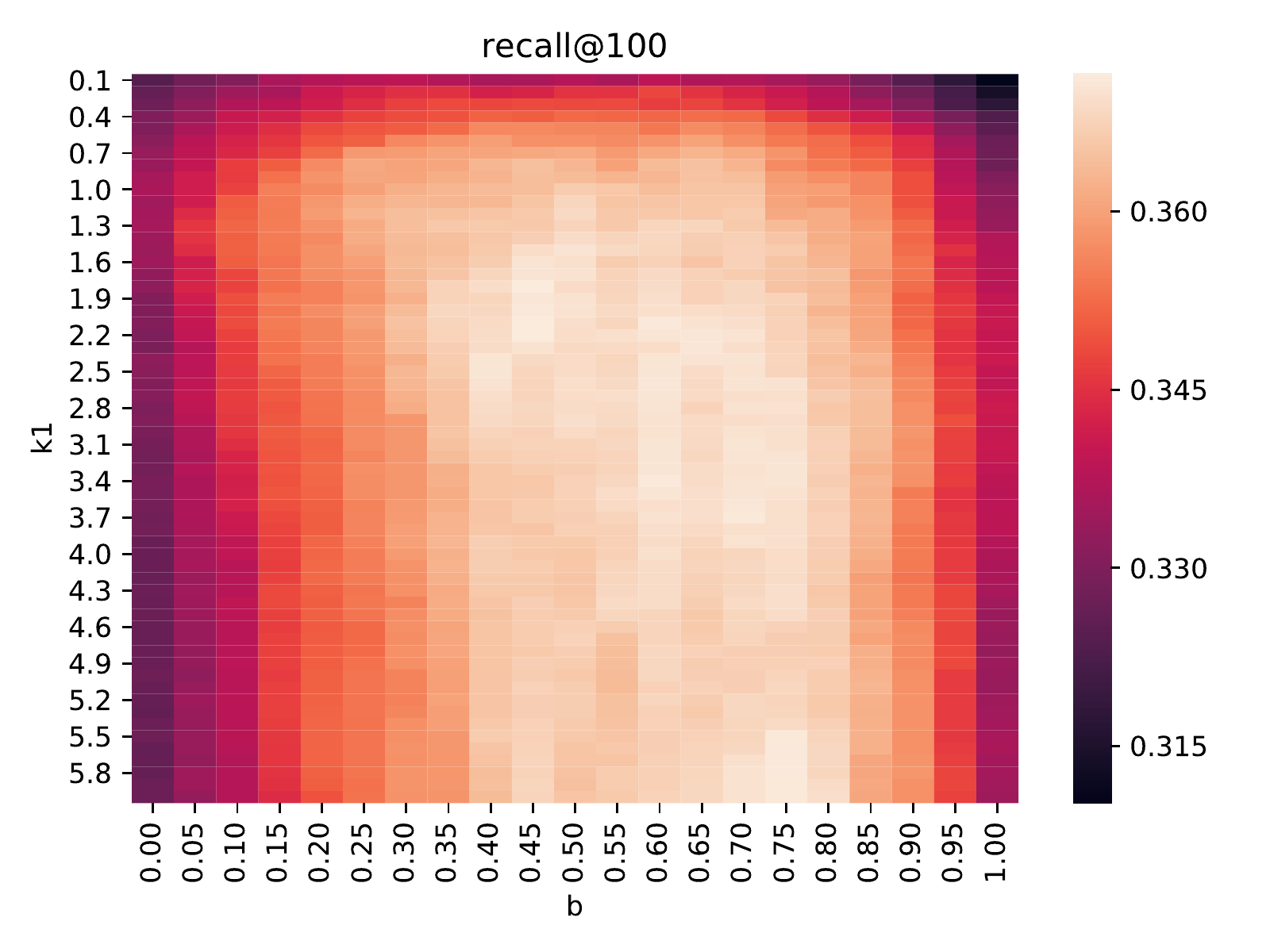}
\includegraphics[scale=0.315]{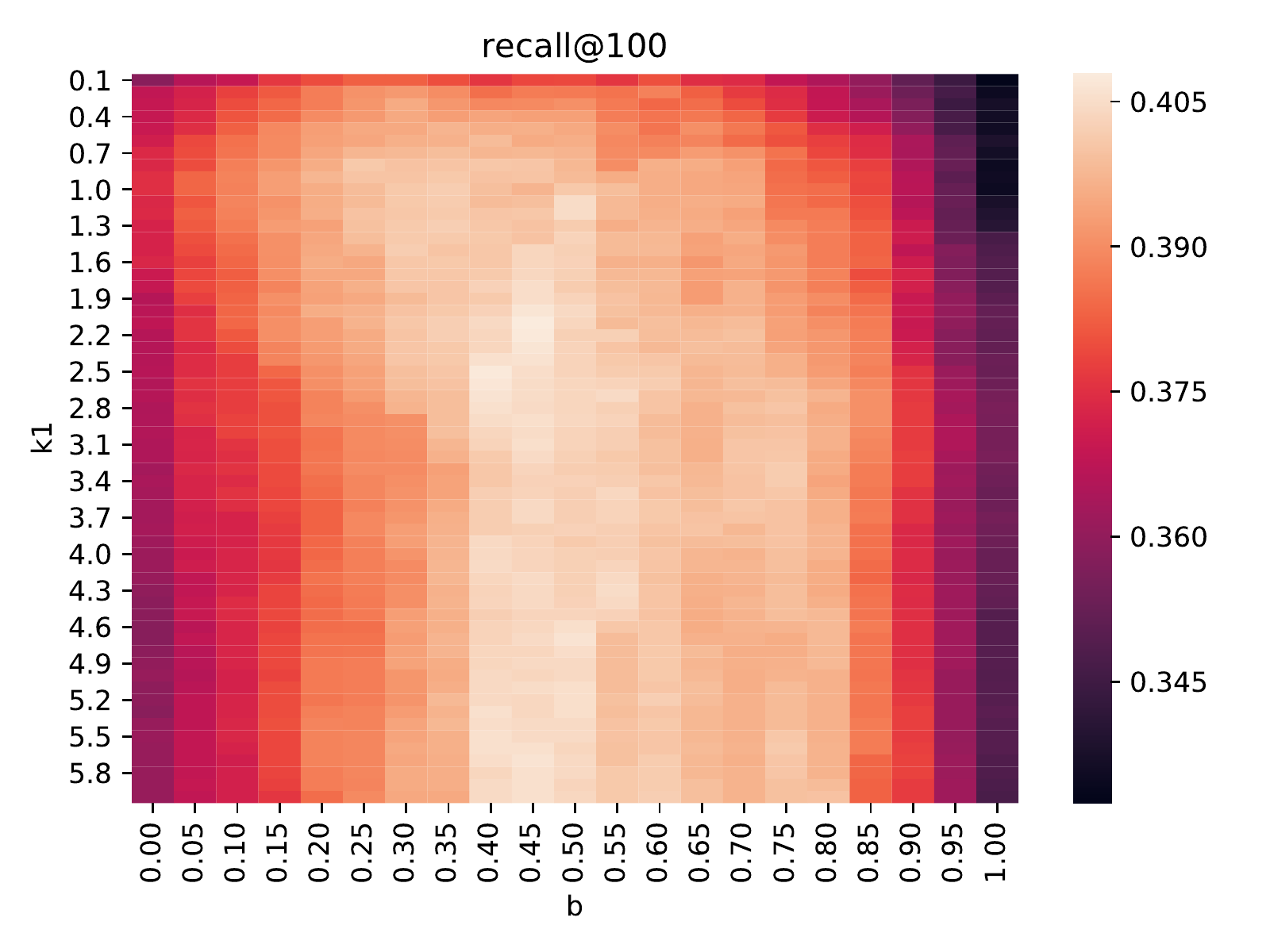}
\caption{Comparison of grid search heatmaps for BM25 using the topic's query over article full text with (a) our relevance judgments, (b) the full set of official judgments, and (c) the set of official relevance judgments filtered to only the topics we assessed. The x-axis sweeps $b\in[0,1]$ and the y-axis sweeps $k_1\in[0.1,6.0]$, and each cell represents the recall@100.}
\label{fig:bm25}
\end{figure*}

We now evaluate our system in an environment that does not utilize human intervention, hyperparameter tuning, or relevance judgements of any sort. This represents a full domain transfer setting. Our pipeline consists of (run tag: run2):
\begin{enumerate}
\item Initial retrieval using untuned BM25 (default parameters, $k_1=0.9$, $b=0.4$), utilizing the question text over the title and abstract of a article. (No date filtering.)
\item Re-ranking using a Vanilla SciBERT model trained on a medical-related subset of MS-MARCO training data. The topic's question field is scored over the article's title and abstract.
\end{enumerate}
The purpose of leveraging the medical-related subset of MS-MARCO is to reduce the risk of domain shift. To produce this subset, we use the MedSyn lexicon~\cite{Yates2013ADRTraceDE}, which includes layperson terminology for various medical conditions. Only queries that contain terms from the lexicon are considered in this dataset, leaving 78,895 of the original 808,531 training queries (9.7\%).\footnote{Several common terms were manually excluded to increase the precision of the filter, such as gas, card, bing, died, map, and fall. This does not qualify as manual tuning because these decisions were made only in consideration of the MS-MARCO training queries, not any TREC-COVID topics.} A list of the query IDs corresponding to this filtered set is available.\footnote{\url{https://github.com/Georgetown-IR-Lab/covid-neural-ir/med-msmarco-train.txt}}

We observe that our automatic \sys run performs highly competitively among other automatic submissions to the TREC-COVID shared task. Although the highest-scoring system in terms of nDCG@10 utilizes a traditional method, we observe that it falls short of neural (e.g., \sys, IRIT\_marked\_base, CSIROmedNIR) in terms of P@5 for fully-relevant articles and the difference between the result are not statistically significant. Furthermore, due to the 88\% and 76\% judged@5 of \sys and CSIROmedNIR, the actual P@5 scores for these systems may very well be higher. Curiously, however, other neural approaches that are generally high-performing (e.g., those used by~\citet{Zhang2020RapidlyDA}) did not rank in the top 10 runs. We do observe that other traditional approaches, such as those that perform query expansion (e.g., udel\_fang\_run3, and uogTrDPH\_QE) also perform competitively in the automatic setting.

\subsection{Analysis}\label{sec:hypertune}

\newcommand{\quest}{Question}
\newcommand{\query}{Query}
\newcommand{\narra}{Narrative}
\newcommand{\chkmk}{\checkmark}
\newcommand{\nochk}{}

\begin{table*}
\centering
\small
\begin{tabular}{@{}llllcrrrrc@{}}
\toprule
\multicolumn{2}{c}{First-Stage} & \multicolumn{2}{c}{Re-Rank} & Filter \\
\cmidrule(r){1-2} \cmidrule(lr){3-4}
Query & Document & Query & Document & 2020 & nDCG@10 & P@5 & P@5 (Rel.) & judged@5 \\
\midrule

\quest & Full text & \quest & Title+abstract & \nochk &     0.7333 &     0.6142 &     0.5467 & 90\% \\
\query & Full text & \query & Title+abstract & \nochk &     0.4190 &     0.5067 &     0.3867 & 70\% \\
\query & Full text & \quest & Title+abstract & \nochk &     0.6244 &     0.7333 &     0.5667 & 94\% \\
\query & Full text & \narra & Title+abstract & \nochk &     0.6133 &     0.5089 &     0.4600 & 82\% \\
\quest & Full text & \quest & Title+abstract & \chkmk & \bf 0.7733 &     0.6774 &     0.6333 & 91\% \\
\query & Full text & \query & Title+abstract & \chkmk &     0.5131 &     0.6267 &     0.4933 & 77\% \\
\query & Full text & \quest & Title+abstract & \chkmk &     0.6844 & \bf 0.7933 & \bf 0.6533 & 100\% & * \\
\query & Full text & \narra & Title+abstract & \chkmk &     0.4898 &     0.5867 &     0.4733 & 70\% \\

\bottomrule
\end{tabular}
\caption{Performance of our system using various sources for the first-stage query text, re-ranking query text, and date filtering. Our official submission is marked with *.}
\label{tab:ablation}
\end{table*}

\paragraph{Initial retrieval parameters}
We now evaluate the hyperparameter tuning process conducted. We first test the following first-stage ranking functions and tune for recall@100 using our judgments: BM25 ($k_1\in[0.1,6.0]$ by 0.1, $b\in[0,1]$ by 0.05), RM3 query expansion~\cite{Jaleel2004UMassAT} over default BM25 parameters (feedback terms and feedback docs $\in[1,20]$ by 1), QL Sequential Dependency Model (SDM~\cite{Metzler2005AMR}, term, ordered, and un-ordered weights by 0.05). Each of these models is tested using with the query or question topic field, and over the article full text, or just the title and abstract. We find that using BM25 with $k_1=3.9$ and $b=0.55$, the topic's query field, and the article's full text to yield the highest recall. We compare the heatmap of this setting using our judgments, the full set of official judgments, and the set of official judgments filtered to only the topics we judged in Figure~\ref{fig:bm25}. Although the precise values for the optimal parameter settings differ, the shapes are similar suggesting that the hyperparameter choices generalize.

\paragraph{Topic fields and date filtering}
Important hyperparmeters of our system include which topic field (question, query, or narrative) to use in which stage, and whether to perform date filtering. We present a study of the effects of these parameters in Table~\ref{tab:ablation}. First, we observe that the filtering of articles to only those published after January 1, 2020 always improves the ranking effectiveness (as compared to models that retrieved from all articles). Indeed, we find that only 19\% of judged documents from prior to 2020 were considered relevant (with only 7\% fully relevant). Meanwhile, 32\% of judged documents after 2020 were considered relevant (19\% fully relevant). We note that although this filter seems to be effective, it will ultimately limit the recall of the system. This observation underscores the value of including a user-configurable date filter in COVID-19 search engines.

We also observe in Table~\ref{tab:ablation} that both first-stage ranking and re-ranking based on the question field may be more effective than using the query field for first-stage ranking and the question for re-ranking. Considering that the nDCG@10 already outperforms the performance of our official submission, and P@5 (fully relevant only) is not far behind with only 91\% of the top documents judged, we can expect that this is likely a better setting going forward. It also simplifies the pipeline and reflects a more realistic search environment in which the user simply enters a natural language question. However, this approach underperforms at identifying partially relevant documents, given by its much lower P@5. In an environment in which recall is important (such as systematic review), the hybrid query-question approach may be preferable. Interestingly, we find that the narrative field usually reduces ranking effectiveness compared to the other settings. This may be due to a large distance between the natural-language questions seen during training and the longer-form text seen at evaluation time. 

\begin{figure}
\centering
\includegraphics[scale=0.8]{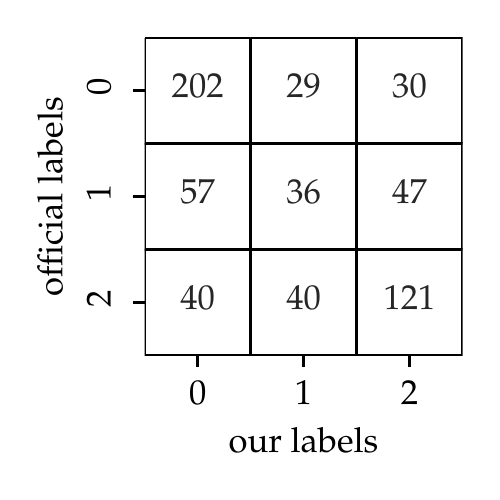}
\caption{Confusion matrix between our non-expert annotations and the official expert TREC labels.}
\label{fig:conf}
\vspace{-1.5em}
\end{figure}

\paragraph{Non-expert judgements}
We found that our non-expert relevance labels did not align well with the official labels; there was only a 60\% agreement rate among the overlapping labels. In 18\% of cases, our labels rated the document as more relevant than the official label; in 23\% of cases ours was rated less relevant. A full confusion matrix is shown in Figure~\ref{fig:conf}.

\begin{figure*}
\centering
\includegraphics[scale=0.6]{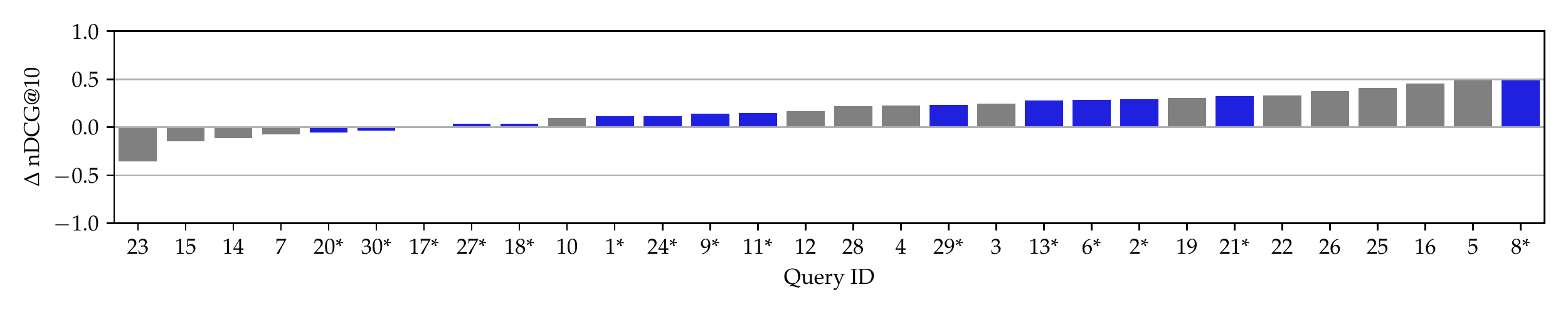}
\includegraphics[scale=0.6]{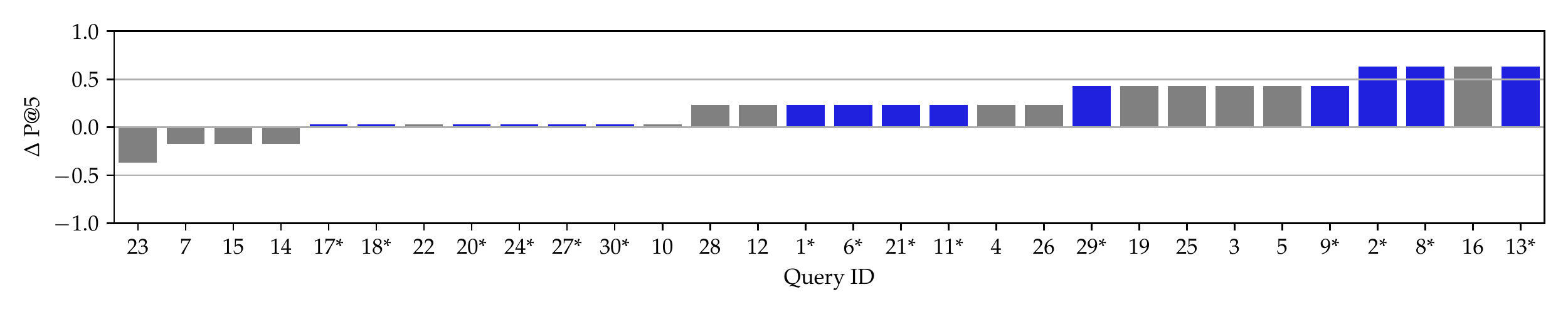}
\includegraphics[scale=0.6]{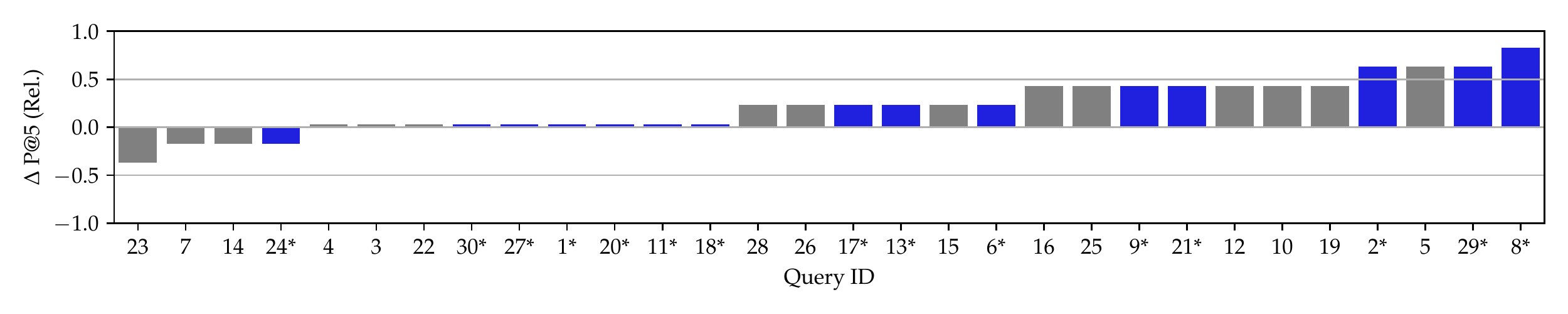}
\caption{Difference in ranking effectiveness between our system and an untuned BM25 model by query for nDCG@10, P@5, and P@5 (fully relevant only). Queries in blue and marked with * were annotated by non-experts and used for hyperparameter tuning.}
\label{fig:deltas}
\end{figure*}

Despite the low agreement rates, the use of domain transfer, and only leveraging the non-expert labels for hyperparameter tuning suggest that it would be difficult to overfit to the test collection. We further investigate whether the subset of queries we evaluated gained a substantial advantage. To this end, we plot the difference in the evaluation metrics between our system and an untuned BM25 ranker in Figure~\ref{fig:deltas}. As demonstrated by the figure, there was no strong preference of our model towards queries that were annotated (marked with * and in blue). In fact, 9 of the 15 highest-performing queries were not in the annotated set (in terms of $\Delta$ nDCG@10). This suggests that our approach did not overfit to signals provided by the non-expert assessments, and that our trained ranker is generally applicable.

\paragraph{Failure cases}
Although our system generally outperforms BM25 ranking, it substantially underperforms for Query 23 (\textit{coronavirus hypertension}). When observing the failure cases, we found that the BM25 model successfully exploited term repetition to identify its top documents as relevant. Meanwhile, our system ranked documents with incidental mentions of \textit{hypertension} highly. This suggests that more effective utilization of approaches that include a count-based component in the ranking score (such as TK~\cite{Hofsttter2020InterpretableT} or CEDR-KNRM~\cite{MacAvaney2019CEDRCE}) could yield improvements.

\section{Conclusion}
In this work we present \sys, a baseline for literature search related to COVID-19. \sys is a two stage approach consisting of an initial BM25 ranker followed by SciBERT-based reranker, a domain specific pretrained language model. \sys is trained on the general domain MS-MARCO passage ranking dataset and evaluated on TREC COIVD-search benchmark in a zero-shot transfer setting. \sys tops the leaderboard among the initial round submissions from 55 teams to TREC-COVID Search shared task, demonstrating its effectiveness.

Through our analysis we find that the parameters for initial retrieval are fairly robust. We also find that recent articles (i.e., those published in 2020) tend to exhibit higher relevance, suggesting the importance of filtering by date for high-precision retrieval. We also find that our non-expert annotation phase helped converge on good hyperparameters, while not likely contributing to substantial overfitting to the test set. Finally, through failure case analysis, we find that count-based approaches may be a good direction to explore in subsequent rounds of the shared task.

\section*{Acknowledgments}
This work was partially supported by the ARCS Foundation.

\bibliographystyle{acl_natbib}
\bibliography{biblio}

\end{document}